\documentstyle[epsfig,here,12pt]{article}
\begin{document}
\newcommand {\ber} {\begin{eqnarray*}}
\newcommand {\eer} {\end{eqnarray*}}
\newcommand {\bea} {\begin{eqnarray}}
\newcommand {\eea} {\end{eqnarray}}
\newcommand {\beq} {\begin{equation}}
\newcommand {\eeq} {\end{equation}}
\newcommand {\state} [1] {\mid \! \! {#1} \rangle}
\newcommand {\sym} {$SY\! M_2\ $}
\newcommand {\eqref} [1] {(\ref {#1})}
\newcommand{\preprint}[1]{\begin{table}[t] %%
           \begin{flushright}               %%
           \begin{large}{#1}\end{large}     %%
           \end{flushright}                 %%
           \end{table}}                     %%
\def\Acknowledgements{\bigskip  \bigskip {\begin{center} \begin{large}
             \bf ACKNOWLEDGMENTS \end{large}\end{center}}}

\newcommand{\half} {{1\over {\sqrt2}}}
\newcommand{\dx} {\partial _1}
\newcommand{\at} {\tilde \alpha}
%%%%%%%%%%
\def\hslash{\not{\hbox{\kern+1.5pt h}}}
\def\Dslash{\not{\hbox{\kern-4pt $D$}}}
\def\cmp#1{{\it Comm. Math. Phys.} {\bf #1}}
\def\cqg#1{{\it Class. Quantum Grav.} {\bf #1}}
\def\pl#1{{\it Phys. Lett.} {\bf #1B}}
\def\prl#1{{\it Phys. Rev. Lett.} {\bf #1}}
\def\prd#1{{\it Phys. Rev.} {\bf D#1}}
\def\prr#1{{\it Phys. Rev.} {\bf #1}}
\def\pr#1{{\it Phys. Rept.} {\bf #1}}
\def\np#1{{\it Nucl. Phys.} {\bf B#1}}
\def\ncim#1{{\it Nuovo Cimento} {\bf #1}}
\def\lnc#1{{\it Lett. Nuovo Cim.} {\bf #1}}
\def\jmath#1{{\it J. Math. Phys.} {\bf #1}}
 \def\mpl#1{{\it Mod. Phys. Lett.}{\bf A#1}}
\def\jmp#1{{\it J. Mod. Phys.}{\bf A#1}}
\def\aop#1{{\it Ann. Phys.} {\bf #1}}
\def\mycomm#1{\hfill\break{\tt #1}\hfill\break}

\begin{titlepage}
\rightline{TAUP-2500-98}
\rightline{WIS-98/13/May-DPP}
\rightline{\today}
\vskip 1cm
\centerline{{\Large \bf The spectrum of multi-flavor $QCD_2$ and}}
\centerline{{\Large \bf the non-Abelian Schwinger equation}}
\vskip 1cm
\centerline{A. Armoni$^a$
\footnote{e-mail: armoni@post.tau.ac.il}
, Y. Frishman$^b$
\footnote{e-mail: fnfrishm@wicc.weizmann.ac.il}
, J. Sonnenschein$^a$
\footnote{e-mail: cobi@post.tau.ac.il} 
and U. Trittmann$^b$
\footnote{e-mail: trittman@wicc.weizmann.ac.il}}   
\vskip 1cm

\begin{center}
{\em $^a$School of Physics and Astronomy
\\Beverly and Raymond Sackler Faculty of Exact Sciences
\\Tel Aviv University, Ramat Aviv, 69978, Israel
\\and
\\ $^b$Department of Particle Physics
\\Weizmann Institute of Science
\\76100 Rehovot, Israel} 
\end{center}
\vskip 1cm
\begin{abstract}
 Massless $QCD_2$ is dominated by classical configurations in the
 large $N_f$ limit. We use this observation to study the theory by
 finding solutions to equations of motion, which are the non-Abelian
 generalization of the Schwinger equation. We find that the spectrum
 consists of massive mesons with $M^2={e^2 N_f\over 2\pi}$ 
 which correspond to Abelian solutions. We generalize previously discovered
non-Abelian solutions  and discuss their interpretation.
We prove a no-go theorem ruling out the existence of  
 soliton solutions. Thus the semi-classical
 approximation shows no baryons in the case of massless quarks, a result
 derived before in the strong-coupling limit only.
\end{abstract}
\end{titlepage}

\section{Introduction}
Since the seminal work of  't Hooft\cite{thooft}, two-dimensional QCD 
($QCD_2$) serves us
as a theoretical laboratory of the real world four-dimensional
QCD. Questions concerning the IR nature of QCD, such as the spectrum
and confinement can be answered in this framework (For recent reviews
see \cite{FSreview,abdalla}).

During the past years, there has been a lot of progress in this field. Apart
from 't Hooft large $N_c$ limit, some other limits were
investigated. In particular, the model with adjoint matter attracted a
lot of attention lately\cite{dalley,gyan,kutasov,AP,FHZ}. This model, due to universality\cite{KS},
is equivalent, in the massless case, to a model with fundamental
fermions and $N_f=N_c$. Another interesting limit which was applied 
in the past, is the massive strongly coupled model ($e\gg m$), in which
the low lying hadronic spectrum was found\cite{DFS}.
The question of confinement versus screening in massless and massive $QCD_2$
has been also thoroughly investigated in recent
years\cite{confscr,FS,AS2,AFS}.

In this paper we discuss the problem of massless $QCD_2$ with large
number of flavors. It seems natural to analyze the theory in this
limit since large N ($N_c$ or $N_f$) simplifies the problem. The space
of massless theories, in the $N_f,N_c$ plane is described in Fig.1

\begin{figure}[H]
  \begin{center}
\mbox{\kern-0.5cm
\epsfig{file=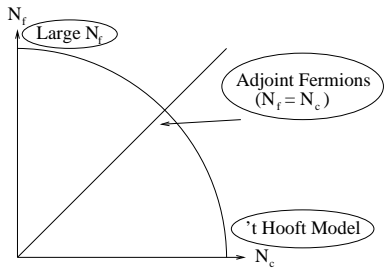,width=5.0true cm,angle=0}}
\label{fc_fig}
  \end{center}
\caption{Massless $QCD_2$ - The flavor-color plane} 
\end{figure}

The diagonal line represents theories with $N_f=N_c$, or adjoint
fermions. The arc represents theories with large $N$. The flow along
this arc seems to be interesting. Starting from small $N_f$ and large
$N_c$ (with $e^2 N_c$ fixed), the spectrum of the theory consists of
a single Regge trajectory. Moving along the arc, we arrive to the
adjoint fermions model, where the spectrum seems to contain infinitely
many Regge trajectories.  We claim that when we approach the large
$N_f$ regime, all these trajectories collapse to a few massive mesons.
Demonstrating this phenomena is the purpose of this paper.

 The main idea that is used in the present paper 
 is that in the bosonized
description of massless $QCD_2$ the color degrees of freedom can be separated 
from the flavor ones, and 
 the ``colored" action is multiplied by an overall factor $N_f$.  Thus, $1/N_f$
plays the role of $\hslash$ and in the limit of large $N_f$ the system
is classical and therefore the physical states are determined by solutions
of the equations of motion\cite{FS}.

 The study of  the equations of motion  transformed into a form of ``non-Abelian
Schwinger equation"\cite{FS}. We investigate three classes of solutions of
this equation: (i) Abelian plane wave solutions, 
(ii) non-Abelian plane wave solutions, and (iii) soliton solutions.
We show that the first class corresponds to mesons with mass of  $M^2={e^2
N_f\over 2\pi}$ and a degeneracy of $N_c^2-1$. The $SU(2)$ non-Abelian solutions
are generalized to $SU(N_c)$ solutions and are shown to relate (on the real
line) to constant solutions. As for soliton solutions, in 
\cite{FS} it was argued that there are no such solutions. Here we present
a proof of this no-go theorem.

The paper is organized as follows: In the second section we describe
the model and derive the equations of motion. In the third section we
show that the equations of motion, in a particular gauge, can be
derived from a Hamiltonian. The trivial Abelian solutions are also
described in this section. Next, in the fourth and fifth section, we
describe non-Abelian solutions to the equations of motion. In the
sixth section, we show that apart from the Abelian solution and
a constant solution (space time independent), there are no other solutions which
can be interpreted as particles in the quantum theory. We arrive to this
conclusion by showing that there are no solitons solutions. Because any
solution which describes a particle with well defined momentum can be
Lorentz transformed into a particle at rest with a time
independent energy-momentum tensor, no solitons implies that there
are no other single particle states in this theory. 

Next, we
show how mass term can affect the spectrum of the theory. The last
section is devoted to a summary.

We use the following conventions: $x^\pm = \half (x^0 \pm x^1)$ and the
Lie algebra generators are normalized such that $tr\ T^a T^b = {1\over
  2}\delta ^{ab}$.

\section{Multi-flavor $QCD_2$}
Massless multi-flavor $QCD_2$ with fermions in the fundamental
representation of $SU(N_c)$
 is described by the following action
\beq
\label{qcd} S=\int d^2x \ tr (-{1\over 2e^2}F_{\mu\nu}^2+i\bar \Psi\Dslash\Psi )
 \eeq
where $\Psi = \Psi ^i_a$, $i=1 \dots N_c$, $a=1 \dots N_f$.

It is natural to bosonize this system, since bosonization in the 
$SU(N_c)\times SU(N_f)\times U_B(1)$ scheme decouples
color and flavor degrees of freedom (in the massless case).
The bosonized form of the action of this theory is given by\cite{FSreview}
\bea
\label{bosonized} \lefteqn{ S_{bosonized} =} \\
 && N_f S_{WZW}(h) + N_c S_{WZW}(g) -\int d^2x\ tr {1\over {2e^2}} F
 _{\mu \nu} F^{ \mu \nu} + \int d^2x\ {1\over 2}\partial _\mu \phi \partial ^\mu \phi
\nonumber \\
 && -{N_f\over 2\pi} \int d^2 x \ tr (ih^\dagger \partial_+ h A_-
+ih\partial_ - h^\dagger A_+ + A_+ h A_- h^\dagger - A_+ A_-)   \nonumber
\eea
where $h$ is a unitary $N_c \times N_c$ matrix which belongs to the color
gauge group, $g\in SU(N_f)$, $\phi$ is the baryon number and $S_{WZW}$ stands for the
Wess-Zumino-Witten action, which for complex fermions reads
\ber
\lefteqn{S_{WZW}(g)={1\over{8\pi}}\int _\Sigma d^2x \ tr(\partial _\mu
g\partial ^\mu g^{-1}) + } \\
 && {1\over{12\pi}}\int _B d^3y \epsilon ^{ijk} \
 tr(g^{-1}\partial _i g) (g^{-1}\partial _j g)(g^{-1}\partial _k g),
\eer
Since we are interested in the massive spectrum of the theory and 
the flavor degrees of freedom are entirely decoupled from the
system and they are massless, we can put aside the $g$ and $\phi$ fields (There is a residual interaction of the zero modes of the $g,h$ and $\phi$ fields, but it is not crucial to our
discussion\cite{KS}).
 We arrive at the following action
\bea
\lefteqn{S=N_f \{
S_{WZW}(h) -\int d^2x\ tr {1\over 8\pi\at } F
 _{\mu \nu} F^{ \mu \nu} } \nonumber \\  
&&
 - {1\over 2\pi} \int d^2 x \ tr (ih^\dagger \partial_+ h A_-
+ih\partial_ - h^\dagger A_+ + A_+ h A_- h^\dagger - A_+ A_-) \}, \nonumber
\eea
where $\at={e^2 N_f\over 4\pi}$.
Since the action is multiplied by an overall $N_f$ factor, the partition
function is dominated by the classical configurations in the
$N_f\rightarrow \infty$ limit. It is equivalent to the $\hbar
\rightarrow 0$ limit. Thus an investigation of the equations of motion
in this limit, captures the quantum behavior of this theory.

The equations of motion for the dual of the gauge field $F={1\over 2} \epsilon ^{\mu \nu} F_{\mu \nu}$ are\cite{Pat,FHS}
\beq
\label{schwinger}
(D^\mu D_\mu + 2\at)F=0,
\eeq
 where $D_\mu = \partial _\mu - i[A_\mu,.]$.
 This equation is the non-Abelian generalization of the Schwinger equation for
 the massive gluon. Finding solutions to this equation is the main
 goal of this paper.

\section{The Hamiltonian and the Abelian solution}
The equation of motion\eqref{schwinger} can also be derived from an effective Hamiltonian. Starting with the action\eqref{bosonized}, 
fixing the gauge $A_-=0$ and integrating over $A_+$, we obtain the
following action 
\beq
S=N_f \left \{ S_{WZW}(h) + \pi \at \int d^2 x \ tr \ \tilde j^+ {1\over \partial _-
  ^2} \tilde j^+ \right \},
\eeq
Where $\tilde j^+={i\over 2\pi} h\partial_- h^\dagger$.

The light-cone energy of this system is
\beq
P^-=-  {e^2\over 4} \int dx^-  \ tr \ j  {1\over \partial _- ^2}
 j, 
\eeq
where $ j = N_f \tilde j^+=\sqrt 2 \bar \Psi \gamma ^+ \Psi$. 

In order to derive to equation of motion, the canonical commutation
relations should be specified. In this case it is the Kac-Moody
algebra
\beq \label{KM} 
[ j^a(x), j^b(y)]={iN_f\over \pi} \delta^{ab} \partial _- \delta(x-y) +
2if^{abc} j^c(x)\delta(x-y)
\eeq  
The equation of motion are therefore
\beq
\partial _+  j^a(x) = i[H, j^a(x)]=-{e^2 N_f\over 4\pi} {1\over
  \partial_-}  j^a(x) + {e^2\over 2} f^{abc} j^b(x) {1\over
  \partial_-^2} j^c(x) 
\eeq
Substituting the relation $\partial _-^2 A_+ = -{e^2\over 2}  j$ (which is the
equation of motion of the gauge field) we obtain
\beq \label{preeom} 
\partial _+ \partial ^2_- A_+ -i [A_+,\partial ^2_- A_+] + \at
\partial _-A_+ = 0,
\eeq  
where now $[.,.]$ stands for the classical $SU(N_c)$ commutator.
This equation is a total derivative. It is the form of
\eqref{schwinger}, in the gauge $A_-=0$. Integrating this equation
with zero boundary condition (by use of the residual gauge freedom)
 we arrive at
\beq \label{eom} 
\partial _+ \partial _- A_+ -i [A_+,\partial _- A_+] + \at A_+ = 0
\eeq  
The above equation has a simple 'Abelian' solution. An Abelian solution
is a solution in which $A_+$ points in some special direction in the
algebra so that the commutator term in \eqref{eom} vanishes.

 Choosing it to be in the 1 direction, we get
\beq 
A^a = \delta ^{a1} \exp i(k_-x_+ + k_+x_-),
\eeq
with $k_- k_+ = \at$. This solution is the non-Abelian analog of the
Schwinger meson in the Abelian case. We argue that in the large $N_f$
limit, the spectrum of the theory contains physical asymptotic states
which are mesons with $M^2=2\at$. Since we have $N^2_c-1$ possible directions for $A_+$,
this plane wave meson solution appears with a degeneracy of  $N^2_c-1$.
 These states were already formed using a
different approach, in \cite{FHS}. Whether these states are
physical, namely asymptotic, at small $N_f$, has not been settled yet
(see discussion in\cite{FHS}). A recent suggestion \cite{Dalley} is
that they correspond to massive poles in the gluon propagator.
 As a result, massless $QCD_2$ is in a screening phase.
 
Apart from the trivial Abelian solution,
eq.\eqref{eom} admits non-Abelian solutions. In the next section we
review the special case of $SU(2)$.

\section{Non-Abelian solutions in the case of $SU(2)$ and the
  energy-momentum tensor}
A solution of \eqref{eom} in the special case of $SU(2)$ was given in
\cite{FS}. We expand $A_+$ in terms of the $SU(2)$ matrices
$T^0=\sigma _3$ and $T^\pm = \sigma _1 \pm i\sigma _2$.
A solution for $A_+$ is

\bea \label{su2}
\lefteqn{
A_+= {\at - k_-k_+\over k_-} T^0 } \\
&& +\sqrt{{(k_-k_+-\at)\at\over
    4k^2_-}}T^+\exp -i(k_-x_++k_+x_-) \nonumber \\
&& +\sqrt{{(k_-k_+-\at)\at\over
    4k^2_-}}T^-\exp i(k_-x_++k_+x_-) \nonumber
\eea
These solutions are truly non-Abelian, since there is no way to
rotate \eqref{su2} to the abelian solution. It is the non-Abelian
analog of Abelian plane-waves.

Using $\partial ^2 _- A_+ = -i\at h\partial _- h^\dagger$ we
can compute the group element $h$ which correspond to the above solution
\beq
 h=\exp -ik_-x_+(T^0+q(T^++T^-)) \exp i(k_-x_++k_+x_-)T^0,
\eeq
where $q={1\over 2}\sqrt{{k_-k_+\over \at} -1}$.

In order to find the mass of the non-Abelian solutions let us
calculate the energy momentum tensor component. The mass of the states
would be $M^2=2P^+P^-$. 

$T^{++}$ is given by the Sugawara construction
 \beq 
T^{++} = {1 \over N_f+N_c} \pi \ tr :j^+ j^+: \label{Tplusquan} ,
\eeq
where the $N_c$ contribution in the denominator is due to normal
ordering. Since our treatment is classical (and $N_f \gg N_c$), we will
use the classical expression
\beq \label{Tplus}
T^{++} = {\pi \over N_f} \ tr (j^+ j^+) = {N_f \over 4\pi \at ^2} \ tr
(\partial_-^2 A_+ )^2
\eeq
The second relevant component can be read from the action
\beq \label{Tminus}
T^{-+} = N_f \pi \at \ tr ({1\over \partial_- }  j^+)^2 = {N_f \over 4\pi
  \at} \ tr (\partial_-A_+)^2
\eeq 
One can check that the conservation of the energy-momentum tensor follows from the
equation of motion \eqref{eom}.

Substituting the $SU(2)$ solution we obtain
\bea
T^{++} = {1\over 2\pi} N_f k_-^2 {k_+ k_- \over \at} (1-{\at \over k_+
  k_-}) \\
T^{-+} = {1 \over 2\pi} N_f k_+k_- (1-{\at \over k_+
  k_-}) 
\eea
 Since $T^{++}$ and $T^{-+}$ are space independent, 
$P^+= \int dx^- T^{++}$ and $P^-=\int dx^- T^{-+}$
diverge. 
This is also the case in the Abelian solution. 
In the later case we can construct 
wave-packets with finite energy.
For the non-Abelian solution a linear combination is not a solution of
the equations of motion and there is no room for a wave-packet
construction. Therefore the interpretation of this solution is not of a
single particle state, but a coherent state of infinite particles,
like a constant electric field in the real 4D world. 

 However, one may also adopt a DLCQ approach where the light-cone space
coordinate is taken to be discretized. In this case the possible values of
$k_-$ have to obey 
$k_- L= 2\pi n$\cite{FS}. If on top of that one requires that
$h(x^-)=h(x^- +L)$ and hence ${\sqrt {k_- k_+ \over \at}} k_-L = 2\pi m$, then the expression of $M^2=2P^+P^-$ takes the following form 
\beq
  M^2 = 2\at N_f^2 ({m^2 \over n}-n)^2 \label{M2}
\eeq
 In the continuum limit of $L\rightarrow \infty$ for finite $n$ we have
$k_-\rightarrow 0$. However, in the limit of  $k_+\rightarrow \infty$
such that $k_+k_-$ is finite, $M^2$ is finite.
 It will be  interesting to look for such states in the DLCQ studies
which are done at finite $N_f$.

 The states in \eqref{M2} decouple in the large $N_f$ limit, except possibly
those with $m=n$, as they are massless.

\section{$SU(N)$ Non-Abelian solutions}
Let us find $SU(N)$, non-Abelian solutions to equation\eqref{schwinger}. An ansatz for
such solutions is constant $A_+$ and $A_-$.
The equation of motion for such solutions takes the following form 
\beq
-[A_+,[A_-,[A_+,A_-]]]  + \at [A_+,A_-] =0 \label{constant}
\eeq

A convenient choice of generators which spans the $SU(N)$ algebra space is
\beq
(T^i_j)^a_b  = \delta ^i _b \delta ^a_j - {1\over N}\delta ^i_j \delta
^a_b
\eeq
The $N^2$ matrices $T^i_j$ are not linear independent since $\sum_i
T^i_i =0$.
In addition, $T^i_j$ obey the following relations:
\bea
&& tr\ T^i_j = 0 \\
&& tr \ T^i_j T^k_l = \delta^i_l \delta^k_j - {1\over N} \delta ^i_j
\delta ^k_l \\
&& [T^i_j,T^k_l]= \delta ^i_l T^k_j - \delta ^k_j T^i_l
\eea

Recall we can always bring $A_-$ say to diagonal form by a constant
gauge transformation. Let us restrict ourselves here to a special
$A_-$ so that
\bea 
&& A_- = T^1_1 \\
&& A_+ = \alpha ^i_j T^j_i,
\eea
where $\alpha$ is a constant Hermitian matrix. 

The most general solution can be brought to the  form
\beq
 A_+ = \at T^1 _1 + \sum _{i=2} ^N v_i (T^1 _i + T^i _1),
\eeq
by use of $SU(N-1)$ rotation in the directions $i=2,...,N$. $v_i$ are
arbitrary real numbers.

These solutions are directly related to the $SU(2)$ solutions which
were described in the previous section. A gauge transformation along
the $T^1_1$ direction, which depends on $x^-$ only $\tilde U = \exp
ix^- T^1_1$
can be used to eliminate $A_-$. The resulting
$\tilde A_+$ is a solution which can be written in the following way
\beq
\tilde A_+=\tilde U^\dagger (x^-) A_+ \tilde U(x^-) \label{gaugeds}
\eeq
Thus $\tilde A_+$ can be written as a plane-wave, similarly
to \eqref{su2} (The $SU(2)$ solution was given as a rotation
of a constant matrix in \cite{FS}). This is a truly $SU(N)$ plane wave.

The discussion at the end of section 4 regarding divergence of energy on the
line, and the situation in the DLCQ case applies here too.

\section{No-Go theorem: There are no solitons}
In this section we would like to prove that there are no solitons
solutions for eq.\eqref{schwinger}. Let's start with the Abelian
case. For the Schwinger equation, assuming static solution leads to
\beq
(-\partial ^2_1+{e^2\over \pi})F=0,
\eeq
which, clearly, has no static solution with finite energy.

In the non-Abelian case, the problem is harder. For a soliton like
solution, only gauge invariant quantities
should be time independent. Thus $A_\mu$ and $F$ may in principle be
time dependent. Our definition for
solitons is solutions in which the Hamiltonian and momentum densities
are time independent
\bea
&& \partial _0 T^{++} =0 \label{cond1} \\
&& \partial _0 T^{+-} =0 \label{cond2}
\eea
In addition, the total energy of the solution should be finite.

Conditions \eqref{cond1} and \eqref{cond2} together with equations
\eqref{Tplus} and \eqref{Tminus} implies
\bea
&& (\partial _+ + \partial _-) tr \ {1\over 2}(\partial_- F)^2=0
\label{cond21} \\
&& (\partial _+ + \partial _-) tr \ {1\over 2}\at F^2=0 \label{cond22}
\eea
Now, let us assume that solitons solutions exists.

{\bf Lemma 1:} For solitons $tr \ {1\over 2}(\partial_- F)^2=  tr \ {1\over 2}\at F^2$.

{\bf Proof:} Multiplying the equations of motion \eqref{preeom} by
$\partial _- F$ (note that $F=\partial _-A_+$) and taking the trace we
obtain
\beq
tr\ \partial_- F\partial_+ \partial_-F +tr \ \at F\partial_- F =0
\eeq
from which
\beq
\partial_+ tr \ {1\over 2}(\partial_- F)^2 + \partial_- tr \ {1\over
  2}\at F^2 =0
\eeq
Using \eqref{cond21} we obtain 
\beq
tr \ {1\over 2}(\partial_- F)^2 - tr \ {1\over 2}\at F^2 = f(x^+) \label{almost1}
\eeq
Since $f(x^+)$ is $x^-$ independent, we can evaluate its value at
$x^-=\infty$. But the L.H.S. of \eqref{almost1} at infinity is zero,
since the field $F$ should vanish at this point (otherwise, the energy
of the system diverges). Thus $f(x^+)=0$. QED.

{\bf Lemma 2:} Solitons obey $  tr \ {1\over 2} F^2 =  tr \ {1\over
  2}\at A_+^2$.

{\bf Proof:} The proof is similar to the proof of Lemma 1. 
Multiplying \eqref{eom} by $F$ and taking the trace we obtain 
\beq
tr\ F\partial_+ F +tr \ \at A_+\partial _- A_+ =0
\eeq
Using \eqref{cond22} we arrive to
\beq
tr\ {1\over 2} F^2 - tr \ {1\over 2}\at A_+ ^2 = g(x^+)
\label{almost2}
\eeq
Now, let us show that $g(x^+)=0$. The L.H.S. of \eqref{almost2} is
zero at $x^-=\infty $ since both $F$ and $tr \ A_+ ^2$ should vanish at this
point. To see that $tr \ A_+^2 =0$ at infinity we multiply equation\eqref{eom}
by $A_+$ and take the trace. we obtain
\beq
tr\ A_+ \partial_+\partial_- A_+ + tr\ \at A_+ A_+ = 0 
\eeq
Since $F=\partial_- A_+=0$ at $x^-=\infty$ and $A_+$ is bounded, $tr\  A_+ A_+$ is
also zero. Hence $g(x^+)=0$. QED. 

Now, let us look at the quantity
\beq \label{quan}
\int _{-\infty} ^{\infty} dx^- tr \ (\partial_-F +\at A_+)^2 
\eeq
By opening the brackets and integration by parts we arrive to
\bea
&& \int _{-\infty} ^{\infty} dx^- tr \ \left ( (\partial_-F)^2  +\at^2 A_+^2
-2\at \partial _- A_+ F \right )= \\
 && \int _{-\infty} ^{\infty} dx^- tr \ \left ( (\partial_-F)^2  +\at^2 A_+^2
-2 \at F^2 \right )
\eea
But according to Lemma 1 and Lemma 2 this quantity is zero. Now
\eqref{quan} is an integral of positive contributions. In addition, 
 \beq 
 tr \ (\partial_-F +\at A_+)^2 ={1\over 2} \sum _a (\partial _-^2 A_+ + \at
 A_+)^a (\partial _-^2 A_+ + \at A_+)^a 
\eeq
Hence
\beq 
\partial _-^2 A_+^a + \at A_+^a=0
\eeq
The solution of this equation is
\beq 
A_+ = e^{i\sqrt {\at} x^-} M(x^+) +e^{-i\sqrt {\at} x^-} M^\dagger (x^+),
\eeq
where $M(x^+)$ is a matrix. 

From the equation of motion \eqref{eom} we get 
\beq
M= C e^{i\sqrt {\at} x^+},
\eeq
where $C$ is some arbitrary matrix.

Thus the solution is
\beq
A_+ = C e^{i \sqrt {\at} x^0} 
\eeq
which is, obviously, not a soliton (It carries infinite energy).

We conclude that since there are no solitons in this theory, There are
no asymptotic fermionic states. Furthermore, apart from the trivial
Abelian solutions, there are no other solutions which can be
interpreted as single-particle asymptotic states. See also discussion
at the end of section 4.   

\section{Adding a mass term}
In this section we would like to comment on the spectrum of the theory
in the presence of a massive fermions with bare mass $m_q$.
The form of the colored part of the action, in the large $N_f$ limit
is given by (see for example \cite{AS2}) 
\bea
\lefteqn{S=N_f \{
S_{WZW}(h) -\int d^2x\ tr {1\over 8\pi\at } F
 _{\mu \nu} F^{ \mu \nu} } \nonumber \\  
&&
 - {1\over 2\pi} \int d^2 x \ tr (ih^\dagger \partial_+ h A_-
+ih\partial_ - h^\dagger A_+ + A_+ h A_- h^\dagger - A_+ A_-)+
\nonumber \\
&&
 e^{\gamma} \sqrt {\at} m_q \int d^2 x \ tr \ (h+h^\dagger)  \}, \nonumber
\eea
where $\gamma$ is the Euler number. The coefficient of the last term depends on the assumed normal
ordering scale. Fixing the gauge, $A_-=0$ and integrating over $A_+$, we obtain the
following action 
\beq
S=N_f \left \{ S_{WZW}(h) +  \pi \at \int d^2 x \ tr \ j^+ {1\over \partial _-
  ^2} j^+ + e^{\gamma} \sqrt {\at} m_q \int d^2 x \ tr \ (h+h^\dagger) \right \}
,
\eeq
 Using the relation between $h$ and
$j^+$ namely, $j^+ = {i\over 2\pi} h\partial _- h^\dagger$, we can expand $h$
in powers of $j^+$ and ${1\over \partial _-}j^+$ as
\beq
 h=1-{2\pi \over i}{1\over \partial_-}j^+ +{({2\pi \over
     i})}^2{1\over \partial _-}(j^+( {1\over \partial _-}j^+))+ \ldots
\eeq
The dominant contribution to the partition function, in large $N_f$, would be obtained
by small values of the field ${1\over \partial _-}j^+ $ and
therefore 
\bea
\lefteqn{S=} \\
&& 
N_f \left \{ S_{WZW}(h) +  2\pi \at \int d^2 x \ tr \ j^+ {1\over \partial _-
  ^2} j^+ -4\pi^2 e^{\gamma} \sqrt {\at} m_q \int d^2 x \ tr \
({1\over \partial _-}j^+)^2 \right \} \nonumber
\eea
Hence the effect of adding a mass term to the action
translates (approximately) to a redefinition of the coupling constant 
${\at}' = \at + 2\pi e^\gamma \sqrt {\at} m_q$. Thus the theory
includes mesons with $M^2 = 2{\at}'$. Note that in this section
we included only the colored part of the theory. In principle, there
are other sectors of the theory that {\em couple} to the color
sector, but here we focus only at special solutions of the equations
of motion in which the other sectors do not appear. In particular, another
massless state exists. This state is responsible to the confining nature 
of the massive theory. A discussion of this state as a solution of the
equations of motion in the large $N_f$ is given in\cite{AS2}.

Therefore, we cannot conclude that the redefinition of the coupling
constant is the only change, when a mass term is added. The full
theory, in the large $N_f$, contains a richer spectrum than the
massless theory, and including solitons in the color-flavor sector\cite{EFHK}.
Also, by expanding $h$ and keeping only one term, we loose the soliton 
(like in the Sine-Gordon case).
  
\section{Summary} 
In this paper we analyzed multi flavor $QCD_2$ with $N_f \gg N_c$.
 We have shown that there are
Abelian solutions that can be interpreted as mesons with mass $M^2
={e^2 N_f\over 2\pi}$.
 In addition there are truly non-Abelian solutions, that cannot have a
 single particle interpretation (See also discussion at the end of
 section 4). Apart from these solutions, we have shown that there are no
non-Abelian solitonic solutions. This means that the spectrum of the
theory consists of the Abelian mesons only.

This observation was made earlier by a different
approach\cite{AS1}. The Hilbert space of the massless theory can be
spanned by the vacuum state, $\state{0} $ and currents that act on this
state, $\state{\Psi}= tr\ J^n \state{0} $. The spectrum of the theory can be found by
diagonalizing $M^2=2P^+ P^-$. Both $P^+$ and $P^-$ can be expressed in
terms of currents (for details see \cite{KS} and \cite{AS1}). The
algebra which governs this calculation is the Kac-Moody
algebra\eqref{KM} in momentum space 
\beq
[J^a _m, J^b _n] = N_f m\delta ^{ab} \delta _{m+n} +if^{abc} J^c
_{m+n}
\eeq
In the large $N_f$ limit, the dominant contribution will be from the
first term. Thus, the non-Abelian terms are suppressed and the spectrum
should be similar to that of a set of 
$N_c^2-1$   multi-flavor massless Schwinger models.

Two remarks are in order. (i) It is important to emphasize that in spite of the 
fact that $A^a$ carries an  index of the color adjoint representation it 
is a gauge singlet, as we fixed the gauge completely. However, the $N_c^2-1$  degeneracy 
  may  be related  to the screening nature of the massless
theory\cite{confscr}. After all in a confining theory we do not expect to see any
remnant of the number of colors in the  low energy spectrum.
(ii) We do not have conclusive evidence that these states survive 
at finite $N_f$. A BRST analysis of the spectrum\cite{FHS} hinted that they may. On the other hand, numerical studies, in
particular for $N_f=N_c$\cite{dalley}, do not find them.   
It is hard for us to imagine that the states at large $N_f$ will disappear from
the spectrum with smaller number of flavors.
This question deserves further investigation.

We would like to add a remark concerning solving (nonlinear)
differential equations for matrices. Associating such an equation with a 
non-Abelian gauge fiexd Lagrangian enables one to find solutions 
 by  choosing a different gauge for which the equations of motion
are simplified, and gauge transforming their solutions.  
 We have used this procedure by solving the easier equation
\eqref{constant}. Gauge transformation of these solutions \eqref{gaugeds} 
produced solutions to the more elaborated equation \eqref{eom}.

Thus it seems that the large $N$ (where $N$ can be either $N_f$ or
$N_c$) two-dimensional massless gauge theories have a rich and interesting
spectrum. The $N_f=1, N_c\rightarrow \infty$ model (the 't Hooft
model) consists of  a single Regge trajectory. The $N_f\rightarrow
\infty$ model consists of few mesons with $N_c^2-1$ degeneracy. An
interesting limit is the adjoint fermions model which can be viewed a
a model of fundamental fermions with both $N_f \rightarrow \infty,
N_c\rightarrow \infty$ and $N_f = N_c$. This model seems to contain
infinite set of Regge trajectories \cite{kutasov,AS1}. When we decrease the number of
flavors from infinity to one we obtain the 't Hooft model. It is
interesting that when the number of colors is decreased from infinity
to few, all the infinite tower collapses to a few mesons. Since we believe
that such an orbit exists in the space of $(N_f,N_c)$, it would be
interesting to find it, maybe with the aid of a numerical study.

%%%%%%%%%%%%%%%%%%%%%%

\section{Acknowledgments}
We thank  A. Zamolodchikov for illuminating discussions.

The work of J.S. is supported in part by the Israel Science Foundation, the US-Israel Binational
Science Foundation and the Einstein Center for Theoretical Physics at
the Weizmann Institute.

The work of U.T. is supported by a MINERVA fellowship.

\end{document}